\documentclass[12pt,preprint]{aastex}
\shorttitle{Spherical Accretion Shock Instability}
\shortauthors{Blondin and Mezzacappa}

\begin{document}

\title{The Spherical Accretion Shock Instability in the Linear Regime}

\author{John M. Blondin}
\affil{Department of Physics, North Carolina State University, Raleigh, NC 27695-8202}
\email{John\_Blondin@ncsu.edu}
\and

\author{Anthony Mezzacappa}
\affil{Physics Division, Oak Ridge National Laboratory, Oak Ridge, TN 37831-6354}
\email{mezzacappaa@ornl.gov}

\begin{abstract}
We use time-dependent, axisymmetric, hydrodynamic simulations to study the linear
stability of the stalled, spherical accretion shock that arises in the post-bounce
phase of core-collapse supernovae.  We show that this accretion shock is stable
to radial modes, with decay rates and oscillation frequencies in close agreement with the linear stability 
analysis of Houck and Chevalier.   For non-spherical perturbations we find 
that the $l=1$ mode is always unstable for parameters appropriate to core-collapse 
supernovae.  We also find that the $l=2$ mode 
is unstable, but typically has a growth rate smaller than that for $l=1$.  
Furthermore, the $l=1$ mode is the only mode found to transition into a nonlinear
stage in our simulations.  This result provides a possible explanation for the dominance 
of an $l=1$ 'sloshing' mode seen in many two-dimensional simulations of core-collapse supernovae.
\end{abstract}


\section{Introduction}

The modern paradigm for core-collapse supernovae includes a critical phase between stellar core bounce and explosion that is characterized by a stalled accretion shock,  during which time neutrino heating is believed to reenergize, or at least play a critical role in reenergizing, the stalled shock [\citep{bhf95,mezzacappaetal98,rj00,liebendoerferetal01,burasetal03,fw04}]. (For a review, see \cite{mezzacappa05}.)  This phase is expected to last of order a few hundred milliseconds.

The past decade has seen significant interest in the multidimensional dynamics of this post-bounce accretion phase. Most two-dimensional supernova simulations exhibit strong turbulent motions below the stalled accretion shock [\citep{hbc92,mwm93,herantetal94,bhf95,mezzacappaetal98,burasetal03,fw04}]. In the past, this turbulent flow was attributed to convection driven by the intense neutrino flux emerging from the proto-neutron star at the center of the explosion. 

However,  \citet[hereafter Paper~I]{bmd03} showed that the stalled accretion shock {\it itself} may be dynamically unstable.  
\defcitealias{bmd03}{Paper I}
By using steady-state accretion shock models constructed to reflect the conditions in the post-bounce stellar core during the neutrino heating phase (as was shown in \citetalias{bmd03}) but characterized by flat or positive entropy gradients, and as such convectively stable, \citet{bmd03} were able to isolate the dynamical behavior of the post-bounce accretion shock {\it per se}. They found that small nonspherical perturbations to the spherical accretion shock lead to rapid growth of turbulence behind the shock, as well as to rapid growth in the asymmetry of the initially spherical shock. This spherical accretion shock instability, or ``SASI," is dominated by low-order modes, and is independent of any convective instability.  

Clearly, once the shock wave is distorted from spherical symmetry, the non-radial flow beneath it is no longer defined 
solely by neutrino-driven convection. The fluid flow beneath the shock is, at least initially, a complex superposition of 
flows generated by convection and by the SASI-distorted shock. Once the shock wave is distorted, it will deflect radially 
infalling material passing through it, leading to highly nonradial flow beneath it. With time, the fluid flow beneath the shock 
may in fact be determined by the SASI and not by convection.  Instabilities such as neutrino-driven convection may be 
important only at early times in aiding the neutrino heating \citep{herantetal94,bhf95,mezzacappaetal98,burasetal03,fw04} 
and in setting the shock standoff radius while the explosion is initiated. The standoff radius will in turn determine the time scale 
over which the SASI may develop.

\citet{janka01} provides a qualitative description of this post-bounce accretion phase in terms of a simple hydrodynamic model, with the intent of providing an analytic model that can be used to investigate the conditions necessary for a successful supernova shock.  In this picture the post-bounce phase is described by a standing accretion shock with outer core material raining down on the shock at roughly half the free-fall velocity. After traversing the shock, this gas decelerates and gradually settles onto the surface of the nascent neutron star.  The pressure of this post-shock gas is dominated by electron-positron pairs and radiation, and as such can be modeled as a $\gamma=4/3$ gas.  The approximation of an hydrostatic atmosphere immediately below the accretion shock then yields the result that the gas density increases as $r^{-3}$ and the gas pressure increases as $r^{-4}$ with decreasing radius behind the shock.  Deeper within this settling region, the gas pressure becomes dominated by non-relativistic nucleons and the temperature becomes roughly constant due to neutrino emission. The flow below this transition radius can thus be approximated by an isothermal hydrostatic atmosphere. The steady nature of this accretion shock and post-shock  settling solution is maintained by a balance between fresh matter accreting through the standing shock, and dense matter cooling via neutrinos and condensing onto the surface of the nascent neutron star \citep{chev89,janka01}.  

This model of core-collapse supernovae described by \citet{janka01} is similar to the analytic models presented by \citet[hereafter HC]{hc92} to investigate spherical accretion flows onto compact objects.  These latter models assume the flow can be treated as an ideal gas with a single effective adiabatic index, $\gamma$, and a cooling function with a prescribed power-law dependence on the local density and temperature of the gas.  
\defcitealias{hc92}{HC}
Using a linear stability analysis \citetalias{hc92} showed that extended shocks (where the shock radius, $R_s$, is much larger than the stellar radius, $r_*$) are unstable to radial oscillations.  For $\gamma =4/3$ this critical radius was $R_s \sim 20 r_*$ or larger, depending on the cooling parameters. This is larger than is expected in the post-bounce accretion phase of core-collapse supernovae, suggesting that the stalled supernova shock is stable to radial perturbations. \citetalias{hc92} examined one case  with $\gamma =5/3$ for the stability of nonradial modes, and found that only the lowest order nonradial mode ($l=1$ in terms of Legendre polynomials) was unstable.  However, the growth rate for the $l=1$ mode was slower than that for the radial ($l=0$) mode. They did not present results for nonradial modes for any case with $\gamma =4/3$.

The focus of this paper is to develop a deeper understanding of the SASI by focusing on its development in the linear regime and in the transition from linear to nonlinear behavior and to supplant through such analyses the numerical findings in \citetalias{bmd03}. In so doing, ties to and extensions of the linear stability analyses of \citetalias{hc92} can be made as well, which in part serve to validate the findings made with our multidimensional hydrodynamics code. 

We begin by describing the model in Section 2 and the numerical method with which we evolve the time-dependent flow in Section 3.  The results from one-dimensional models are reported in Section 4, along with the corresponding results from \citetalias{hc92}.  In Section 5 we present two-dimensional simulations, quantifying the SASI growth rate as a function of the Legendre wave number, $l$, and illustrating the physical origin of the $l=1$ instability.

\section{Spherical Accretion Shock}

We begin with the model of the post-bounce accretion shock and ``settling"
flow beneath it presented in \citetalias{bmd03}. As we now discuss, this
model is a limiting case of the model presented in \cite{janka01} and is
defined following the prescription outlined in the earlier work by
\citetalias{hc92}. In our model, we assume we have an ideal gas equation of
state, a cooling function, and a hard reflecting boundary at the surface of
the accreting compact object. We make the additional assumption that the gas
in the postshock region is radiation-dominated---that is, that we have a
single adiabatic index. This is appropriate for conditions in the postbounce
stellar core at a time when explosion is initiated, during which time we
expect an extended heating region overlaying a thin cooling region above the
proto-neutron star surface. Of course, the equilibrium radius of the
accretion shock is determined by the magnitude of the cooling in this
cooling layer: stronger cooling leads to a shock radius closer to the inner
reflecting boundary, while weaker cooling results in a shock with a large
stand-off distance from the accreting object. While one could include a
neutrino heating term in such a model \citep{bg93}, this would possibly
introduce convection in the dynamics, complicating the analysis.  Because
our goal is to separate the effects of shock-driven turbulence from
thermally-driven convection, we do not include any heating term in our
models.

The time-evolution of the flow is given by the Euler equations for 
an ideal gas described by a 
velocity, $\bf u$, a mass density, $\rho$, and an isotropic thermal pressure, $p$: 
\begin{equation}
\partial_t \rho + \nabla \cdot \rho {\bf u} = 0
\end{equation}
\begin{equation}
\partial_t \rho {\bf u} + \nabla \cdot \rho {\bf uu} + \nabla p = -\rho{GM}/{r^2}
\end{equation}
\begin{equation}
\partial_t  \rho{\cal E}  + \nabla \cdot (\rho{\cal E}{\bf u} + p{\bf u}) = -{\cal L} 
\end{equation}
where the total energy per gram is given by
${\cal E} = \frac{1}{2}{\bf u}^2 + e - GM/r$, the internal energy, $e$, is given by
the equation of state: $\rho e = p/(\gamma -1)$, and $M$ is the mass of the accreting star.

Following \citetalias{hc92}, the cooling term is parameterized by two power-law exponents:
\begin{equation}
{\cal L} = A \rho^{\beta-\alpha} p^\alpha .
\end{equation}
Using the parameterization of effective neutrino cooling provided by \citet{janka01}, with
$\dot E \propto \rho T^6$, and assuming $P\propto T^4$, we arrive at values of $\alpha = 3/2$ and
$\beta=5/2$ for our model.

\begin{figure}[!hbtp]
\begin{center}
\includegraphics{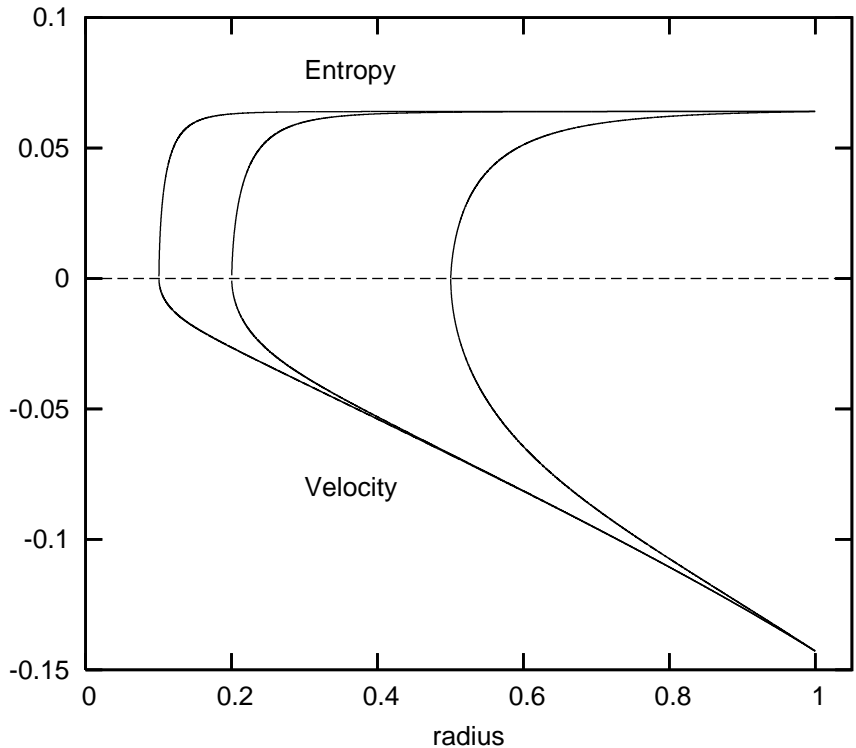}
\caption{Equilibrium solutions for the spherical accretion shock model of
a core-collapse supernova are shown for different values of the radius of
the proto-neutron star ($r_*$ = 0.04, 0.2, 0.5) relative to the shock radius.
The velocity and entropy are integrated inwards from the
accretion shock at $R_s=1$ to the stellar surface at $r_*$.}
\label{fig:analytic}
\end{center}
\end{figure}

Following Paper I, we normalize the problem such that $GM=0.5$, $\dot M=4\pi$, and the
equilibrium accretion shock is at
$R_s=1$.   Note that this results in the same normalization for density, velocity, and
pressure as used in \citetalias{hc92}.  Assuming free-fall velocity just above the accretion
shock, the immediate post-shock values are then given by
\begin{equation}
u_s = \frac{\gamma - 1}{\gamma +1}, 
\rho_s = \frac{\gamma + 1}{\gamma -1}, 
p_s = \frac{2}{\gamma +1}.
\end{equation}
The equilibrium solutions are obtained by integrating the Euler equations inward
from the shock front until $u=0$, corresponding to the surface of the accreting star.
These equilibrium solutions are shown in Figure \ref{fig:analytic} for three different values of
$r_*$, and hence three different stellar radii relative to the normalized shock radius.

The entropy profiles show in Figure \ref{fig:analytic} illustrate the regime of strong
neutrino cooling in this supernova model.  
In the absence of cooling the flow would remain isentropic,
thus the drop in entropy seen in Figure \ref{fig:analytic} reflects the local efficiency of cooling,
which is predominantly confined to a thin layer at
the surface of the accreting star.  Furthermore, as the stellar radius recedes from the shock
front (a more extended shock region), this cooling layer becomes progressively thinner.

There are two characteristics of our models that warrant further discussion,
particularly as they compare to past supernova models: (1) The mass
accretion rate through the shock is assumed to be constant. In reality, the
accretion rate decreases with time owing to the decreasing density and
velocity in the preshock gas. Therefore, the results we present here would
be enhanced if the drop-off in mass accretion rate were included. (2) We
assume a high Mach number for the shock (anything $\sim3$ or larger would be
sufficiently large to be consistent with our models). 
This assumption is consistent with the profiles found in
supernova models (e.g., see \cite{mezzacappaetal01}).

\section{Numerical Model}

As described in \citetalias{bmd03} in more detail, 
we use the time-dependent hydrodynamics code VH-1 
(\url{http://wonka.physics.ncsu.edu/pub/VH-1/}) to study the 
dynamics of a spherical accretion shock in both one and two
dimensions.  A critical aspect of this numerical implementation
is the use of dissipation to maintain a smooth flow in the
absence of any perturbations \citepalias{bmd03}.

These models
are described by four parameters, $\gamma$, $\alpha$, $\beta$,
and $r_*$ (or $A$).  Note that although the parameters $A$ and $r_*$ are
equivalent, they are not related by a closed-form expression. 
We therefore numerically
search for the value of $A$ that produces a steady shock
at a radius of unity for a given inner reflecting boundary at
a radius of $r_*$.
This steady-state solution is then mapped onto a
radial grid extending from $r_*$ to $2 R_s$ to initialize the numerical simulations. 
The shock front is smoothed over two numerical zones to minimize spurious
waves at the start of the simulation.

The numerical resolution required for an accurate solution depends on the 
parameters of the problem, namely the scale-height of the cooling region.
We found appropriate resolutions empirically by evolving one-dimensional
simulations at various resolutions and comparing the time-evolution of the
shock radius.  We used grids with 300 to 450 radial zones, including a 
small increase (never more than 1\%) in the radial width of the zones to
maximize resolution near $r_*$ where rapid cooling generates strong gradients.
The two dimensional simulations 
use the same number of zones used in the radial direction to cover
the polar angle (a limitation of the parallel algorithm used by VH-1)
from $0$ to $\pi$, assuming axisymmetry about the polar axis.
The fluid variables
at the outer boundary are held fixed at values appropriate for
highly supersonic free-fall at a constant mass accretion rate,
consistent with the analytic standing accretion shock model.  Reflecting
boundary conditions were implemented at the inner boundary, $r_*$.
For the two-dimensional simulations,
reflecting boundary conditions were applied at the polar boundaries
and the tangential
velocity was initialized to zero everywhere in the computational 
domain.

\section{Code Verification}

The results of \citetalias{hc92} provide an opportunity to verify our numerical code 
on a time-dependent flow problem of direct relevance to core-collapse supernovae.
Their linear stability analysis shows that spherical accretion shocks are 
unstable to growing oscillations in the shock radius (for the fundamental mode)
if the shock is relatively extended.

To confirm this stability analysis and to verify our numerical code, we have 
run simulations matching the parameterization for post-supernova fallback
used in \citetalias{hc92}.  They considered the
problem of fall back onto the nascent neutron star on the time scale of hours to days
following the supernova explosion.  In this case the gas is optically thick
and radiation-dominated ($p\propto T^4$, $\gamma=4/3$, as in the present post-bounce model).  
Near the surface of the accreting neutron star, the gas is  loosing energy
through neutrino cooling with a negligible density dependence but 
a strong temperature dependence 
($\dot E \propto T^{10}$).   These properties are approximated in the 
present model by considering the parameter values
$\alpha=\beta=2.5$ \citep{chev89}.  We note that the fall-back solutions used by
\citetalias{hc92} and the post-bounce solutions described in this paper are very 
similar.

\begin{figure}[!hbtp]
\begin{center}
\includegraphics{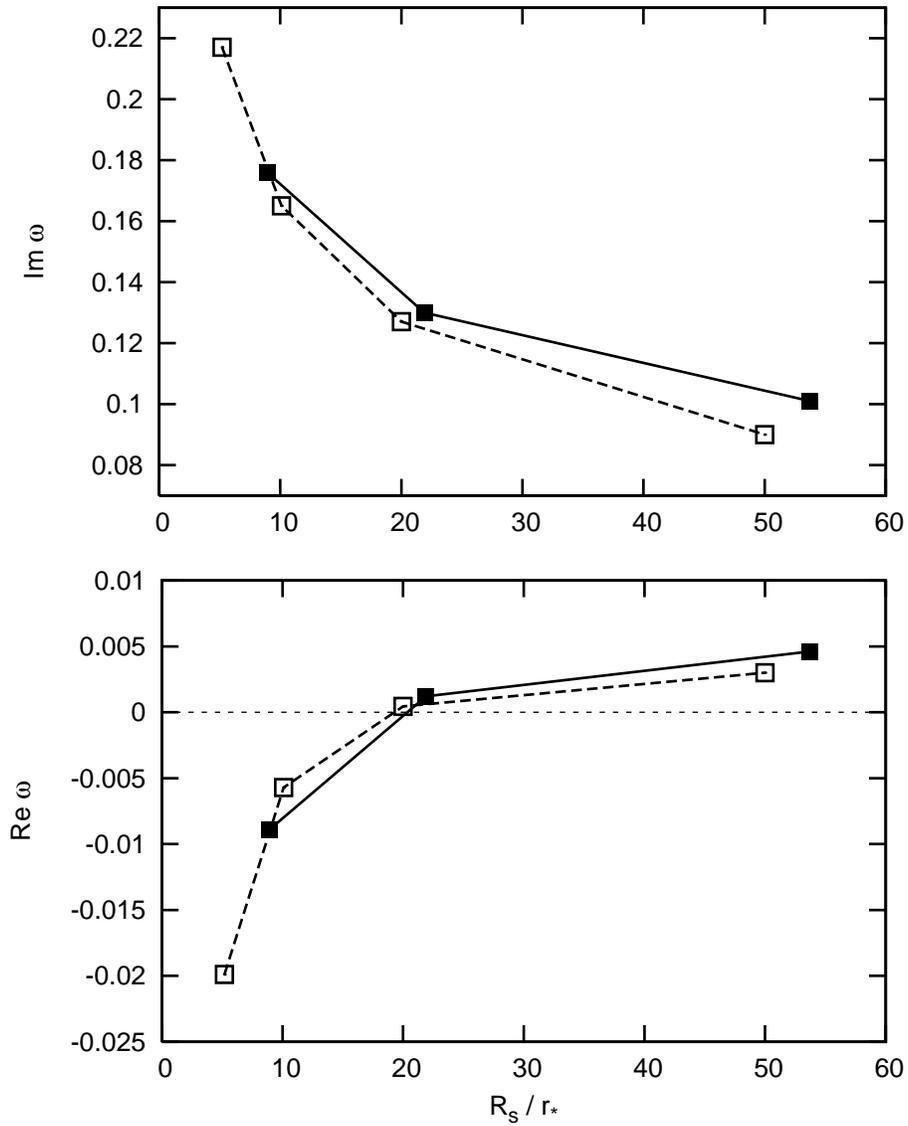}
\caption{The real and imaginary parts of the growth rate, $\omega$, as a function of the shock height, $R_s/r_*$ for the fall-back model.  Results are shown 
for the linear stability analysis of \citetalias{hc92} (solid line) and for the 1D numerical simulations described in this paper (dashed line). Note that the values from \citetalias{hc92} have been
transformed to the units used in this paper. }
\label{fig:omega}
\end{center}
\end{figure}

We perturb the equilibrium solutions by dropping an over-dense shell onto
the shock, which compresses and pressurizes the shock region.  The overpressure
drives the shock back outwards, and in addition a strong pressure wave rebounds 
off the stellar surface and drives the accretion shock out even faster.  This sets up
an oscillation of the shock region on the sound crossing timescale.  For parameters
typical in core-collapse supernovae this oscillation is damped, as it was in the adiabatic 
models of \citetalias{bmd03}.  For very extended shocks this oscillation is overstable;
for the fall-back case this critical radius is $r_*\approx 0.05$.

To extract a complex growth rate from these simulations we fit the simulation
data for the shock radius, $R_s(t)$, to an analytic function of the form
\begin{equation}
R_s(t) = R_0 + R_1e^{\omega_r t}\sin(\omega_i t + \delta).
\end{equation}
Using
least-squares fitting we obtain the real and imaginary parts of the growth rate, $\omega$.
The results from several simulations of the fall-back model 
are shown in Figure \ref{fig:omega}, together with the linear growth rates derived by \citetalias{hc92}.  
There is only a limited range in $R_s/r_*$
where the linear analysis and the numerical simulations overlap, but within that overlap the 
agreement is remarkably good.  

Simulations of very extended shocks ($r_* \ll R_s$)
become computational expensive because of the large dynamic range in spatial coverage and the extremely short time scale.  As $r_*$ becomes 
much smaller than $R_s$, the region of strong cooling (for $\gamma = 4/3$) shrinks to a small
fraction of $r_*$.  This forces the use of a high resolution spatial grid near $r_*$, which in
turn requires a large number of computational zones and a very small time step due to
the Courant condition: $\Delta t <  \Delta r / c_s$.  As an extreme example, to simulate 
a model with $r_* = 0.01$ required 200,000,000 timesteps.

These simulations do two important things: they confirm the linear
stability analysis of \citetalias{hc92}, and they validate our time-dependent
numerical model.

\section{Linear Evolution of the SASI}

We performed a series of two-dimensional axisymmetric simulations following 
the same procedure outlined above for one dimension, down to the same
radial gridding for a given model.  We experimented with a variety of ways to
perturb  the equilibrium solution with the goal of exciting a single mode (in terms
of spherical harmonics) with as little power in other modes as possible.
This goal was best achieved when using density
enhancements in the preshock gas as had been used in 1D.  These density
variations were typically between 0.1\% and 1\%; large enough to excite the instability but 
small enough that the perturbations could grow in amplitude by more than an
order of magnitude while still remaining small.  This extended regime of
linear growth facilitated an accurate measurement of the linear growth rate.
As in \citetalias{bmd03},
all of the two-dimensional simulations were unstable.  Here we attempt to
quantify the growth rate of this instability in the linear regime as a function of the wave number.

To track the importance of various modes affecting the stability of a spherical
accretion shock, the evolution was tracked using Legendre polynomials.  Again, after
trying several methods, we found the best approach was to first integrate the amplitude
in a given harmonic for a fixed radius, 
\begin{equation}
G(r) =   \int A(r,\theta) P_l(\cos\theta) d\cos\theta
\end{equation}
and then integrate the power for that harmonic
over radius:
\begin{equation}
Power = 2\pi\int    \left[G(r)\right]^2 r^2 dr 
\end{equation}
where $A(r,\theta)$ represents some local quantity affected by the perturbed flow,
e.g., entropy or pressure, and $P_l$ is the Legendre polynomial of order $l$.  

\begin{figure}[!hbtp]
\begin{center}
\includegraphics{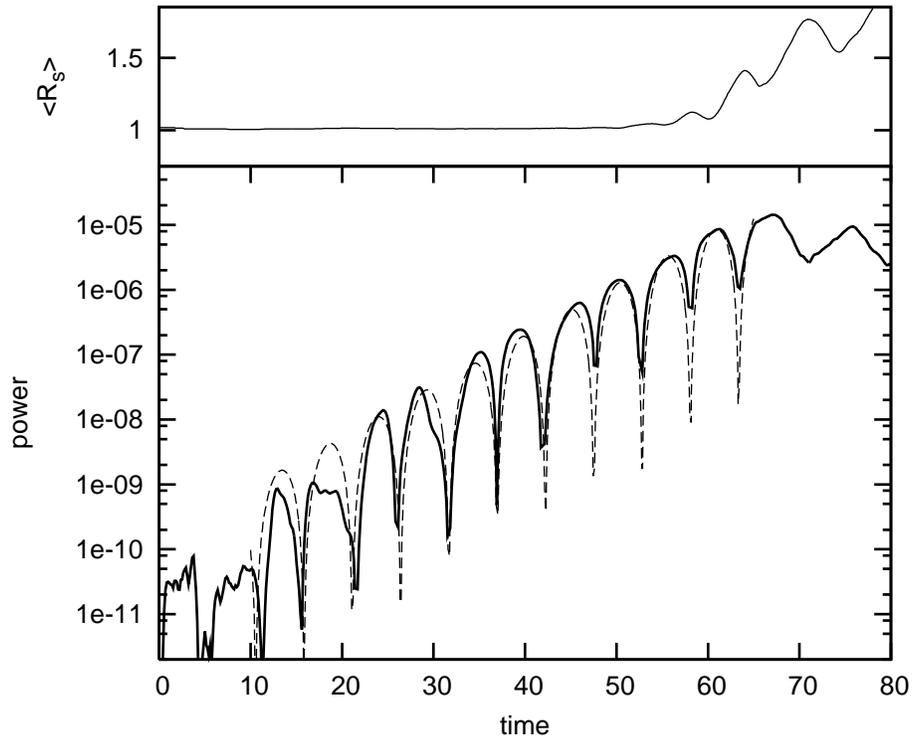}
\caption{The growth of the SASI in a simulation with $r_* = 0.2$ 
is quantified here by the power in the perturbed entropy for the 
$l=1$ mode (solid line).  The best fit to this growth curve is shown as a dashed line.
The SASI becomes nonlinear at a time around $t\approx 60$, as shown by the
deviation of $\langle R_s\rangle$ from unity.}
\label{fig:sasigrowth}
\end{center}
\end{figure}

An example of the linear growth of the SASI is shown in Figure \ref{fig:sasigrowth} for
$r_*=0.2$.  These simulations exhibit a well-defined
regime of exponential growth spanning at least an order of magnitude in amplitude.
The beginning of a simulation is typically marked by a complex pattern of waves, but
given an appropriate initial perturbation, a single mode soon dominates the evolution.
To provide guidance on the relevance of the linear regime, we also show the angle-averaged
shock radius throughout the evolution.  During the linear regime, when perturbations to
the spherical accretion flow are small, we expect the shock to remain nearly stationary.
Once the average shock radius begins to deviate substantially from unity, the SASI has 
entered the non-linear regime.

\begin{figure}[!hbtp]
\begin{center}
\includegraphics{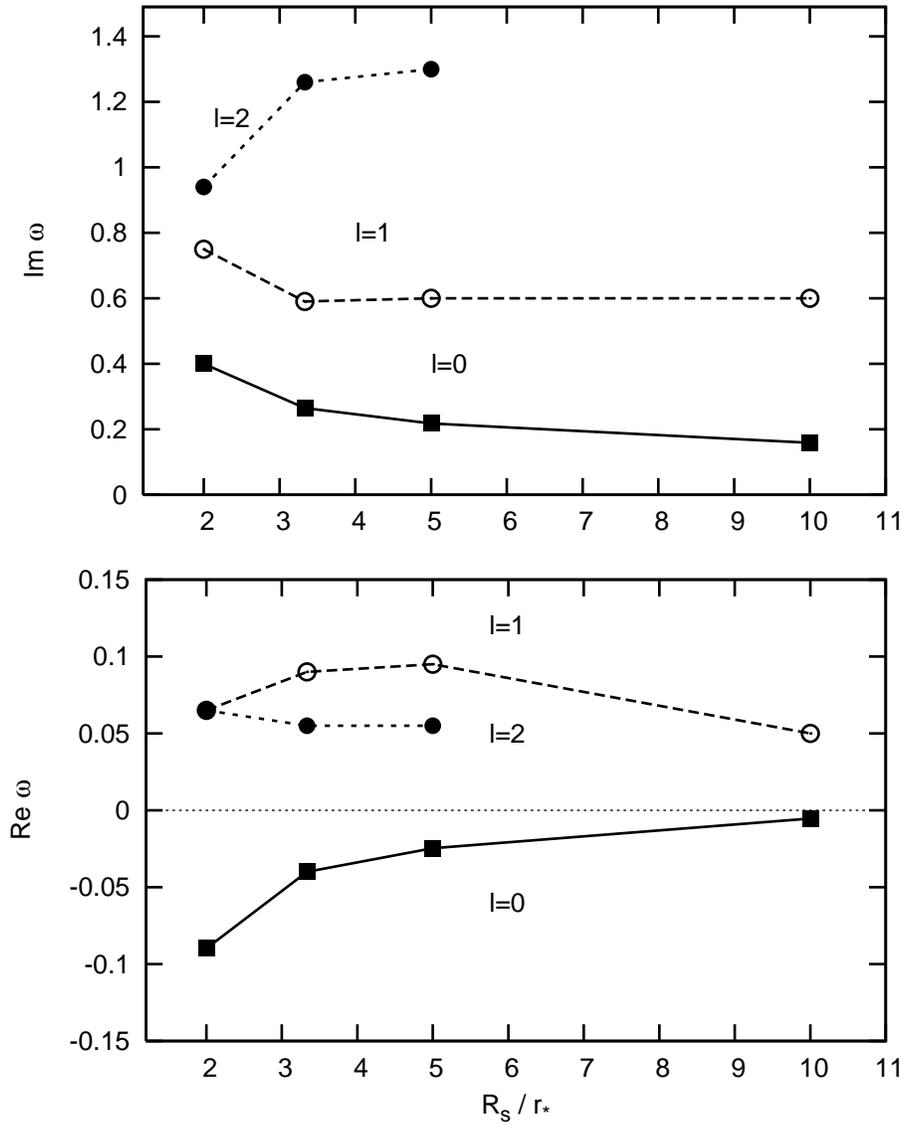}
\caption{The real and imaginary parts of the growth rate, $\omega$, as a function of the shock radius
relative to the stellar radius, $R_s/r_*$ for three different axisymmetric modes, $l=0$, 1, and 2. }
\label{fig:omega2d}
\end{center}
\end{figure}

As in the analysis of the one-dimensional simulations, we can fit these growth curves
from two-dimensional simulations with an exponentially growing sinusoid. In this
case, we are fitting the power, not the radius:
\begin{equation}
F(t) = F_1e^{2\omega_r t}\sin^2(\omega_i t + \delta).
\end{equation}
The fitted frequencies are shown in Figure \ref{fig:omega2d} for four different values 
of $r_*$ (0.5, 0.3, 0.2, and 0.1).  We did not attempt
simulations for smaller values because they would have required an extremely long
integration, and the results for the $r_* = 0.1$ model were sufficiently noisy that we 
did not expect to be able to extract clean growth rates from more extended models.
Note, however, that the growth rate of the $l=1$ mode appears to be decreasing for
very extended shocks and that the $l=0$ mode becomes unstable for $r_* < .05$.

We do not consider values of $r_*$ larger than 0.5 because this would not be consistent 
with our fundamental starting assumption of having conditions near explosion and a 
postshock gas described by a single adiabatic index. Under such conditions, we would 
have a large heating region dominated by radiation and a thin cooling layer at its base. 
A larger value of $r_*$ would imply a larger cooling region and, generally, a postshock 
region composed of gases with different adiabatic indices.

We could not isolate modes with values of $l>2$, nor could we adequately measure the
growth of the $l=2$ mode in the most extended models with $r_*=0.1$.  The results for the spherically
symmetric mode ($l=0$) are taken from the one-dimensional simulations.  As expected,
the frequency of oscillation is a monotonically increasing function of the wavenumber.
The growth rate, however, is not.
In all cases we found  that $l=1$ is the most unstable, and is always unstable. 

\begin{figure}[!hbtp]
\begin{center}
\includegraphics{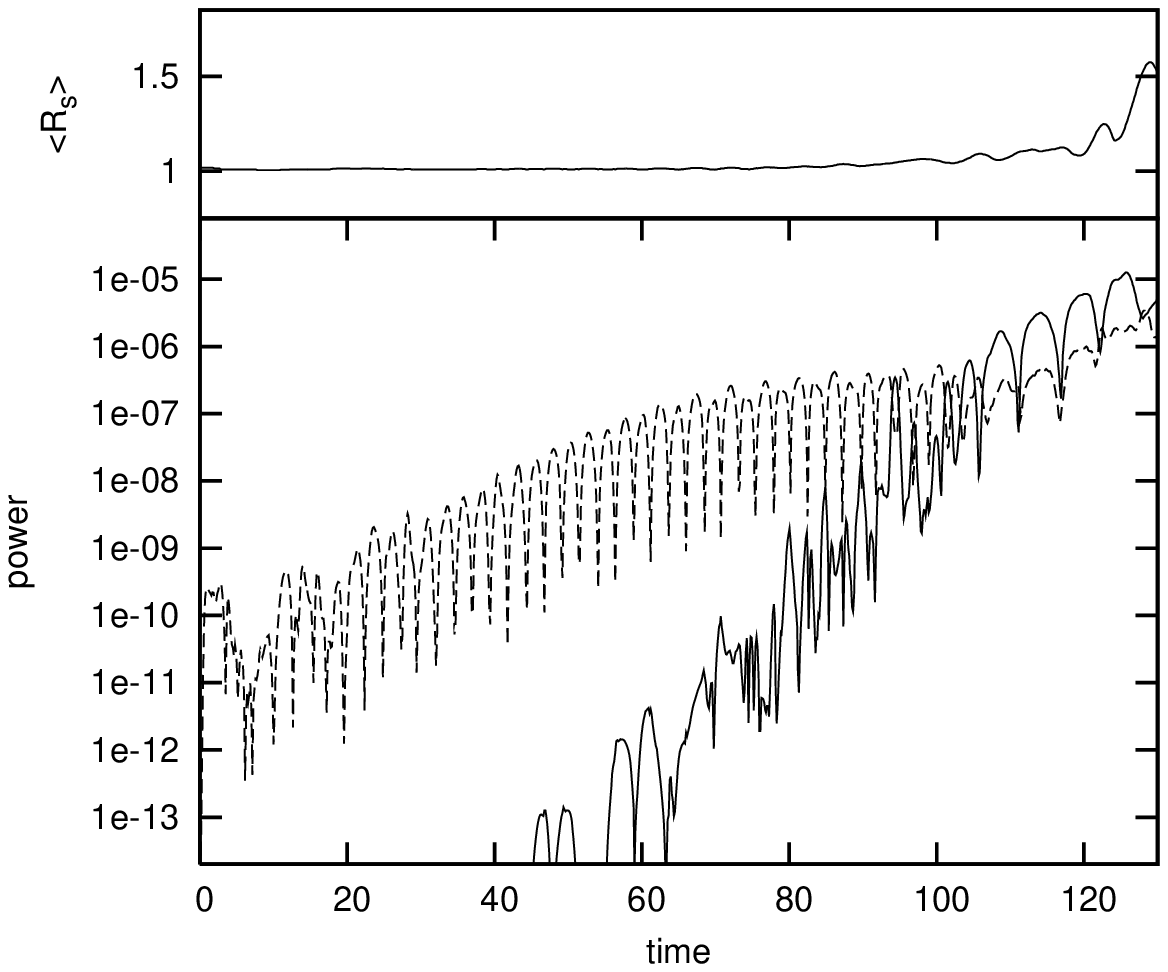}
\caption{The growth of the SASI  in a simulation
excited with the $l=2$ mode (dashed line)  but dominated at late times by the $l=1$ mode (solid line).
The SASI becomes nonlinear at a time around $t\approx 120$, once the $l=1$ mode becomes dominant.}
\label{fig:saturate}
\end{center}
\end{figure}

At late times the evolution is always dominated by the $l=1$ mode.  In fact,
this is the only mode that we have observed to reach a nonlinear stage.  We show in Figure \ref{fig:saturate}
the growth of the different modes in a simulation for which we carefully excited the
$l=2$ mode and not the $l=1$ mode.  While the $l=2$ mode grows substantially over the
course of a dozen oscillations, it stops growing before reaching the nonlinear stage.  
In contrast, the $l=1$ mode grows up out of the noise and becomes nonlinear at a time of 
about $t\approx 120$.  We speculate that the linear growth of the $l=2$ mode stalls because
power in that mode is lost to the rapidly growing $l=1$ mode.  
Note that the power in $l=1$ is very chaotic during this episode when the
$l=2$ growth stalls.
In fact, a comparison of Figures \ref{fig:sasigrowth} and \ref{fig:saturate} shows that 
the overall growth rate for $l=1$ is steeper in this latter model than observed
for a simulation with only an $l=1$ mode excited.  While it might be possible for the $l=2$ mode to
reach a nonlinear stage if the $l=1$ mode was completely suppressed, one would not expect such
an artificial situation to happen in Nature.

To understand the physical origin of the SASI, we first note that the oscillation frequency of the 
SASI corresponds roughly to the time it takes a 
sound wave to cross the spherical accretion shock cavity.  For the model
with $r_*=0.2$, a sound wave travels from $R_s$ to $r_*$ in a time of 1.51.  Neglecting
the path around the stellar surface at $r_*$, a sound wave would travel back and
forth across the spherical cavity in a characteristic time $\tau_s \approx 6$, giving an
oscillation frequency of $\omega_i \approx 1$.  Note that a more realistic path around
the central star would give a longer travel time, both because of the longer path length and
because the sound speed is slower at larger radii.  Thus, one would expect a frequency 
somewhat smaller than unity, in agreement with the results shown in Figure \ref{fig:omega2d}.
In contrast, the advection time for a parcel
of gas to drift in from $R_s$ to $r_*$ is 14.3 for this same model.  Therefore, any
small perturbations in advected quantities (e.g., $u_\theta$, vorticity, entropy) are
seen to drift inwards on a time scale much longer than the characteristic time scale
of the SASI.  

\begin{figure}[!hbtp]
\begin{center}
\includegraphics[width=6.5in]{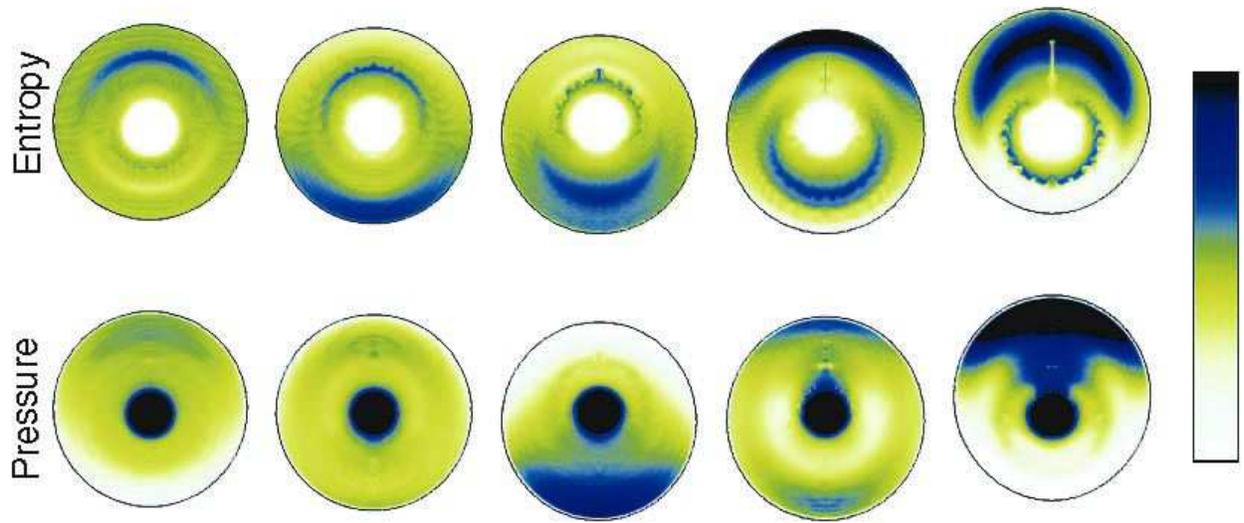}
\caption{An example of the perturbed variables taken from the simulation excited with
an $l=1$ mode.  The entropy is shown in the top row and pressure in the bottom row.
In each case blue represents positive deviations from equilibrium and white represents 
negative.  Time evolves to the right over one period of oscillation; the first and last images 
represent the same phase.  An mpeg animation of this evolution is available on line. }
\label{fig:2dvar}
\end{center}
\end{figure}

This difference between the propagation of sound waves and the slower drift of 
advected perturbations can be seen in Figure \ref{fig:2dvar}, where we show the 
time evolution of two flow quantities.  In the case of entropy, 
which serves as a marker of fluid elements when the gas is adiabatic (which is the case all but near
the accreting surface), the perturbations advect radially inward and fade away into
the low-entropy gas of the cooling layer. These perturbations are generated at the shock and advect inwards 
with the accretion flow.  As such, they have no means of directly 
influencing the accretion shock.  Furthermore, there is no evidence of pressure perturbations
at small radii that might represent acoustic waves originating from these flow perturbations
as they advect inward---i.e., of a vortical--acoustic feedback.  In contrast, the bottom row in Figure \ref{fig:2dvar} shows something that
resembles a standing wave pattern rather than features drifting radially inward with the flow.
For example, the region of high pressure does not propagate across the equator of the
shock, but rather fades in one hemisphere only to grow in the low pressure region in the opposite
hemisphere. Based on these observations, we conclude that the
SASI is the result of a growing standing pressure wave oscillating inside the
cavity of the spherical accretion shock.  

\begin{figure}[!hbtp]
\begin{center}
\includegraphics{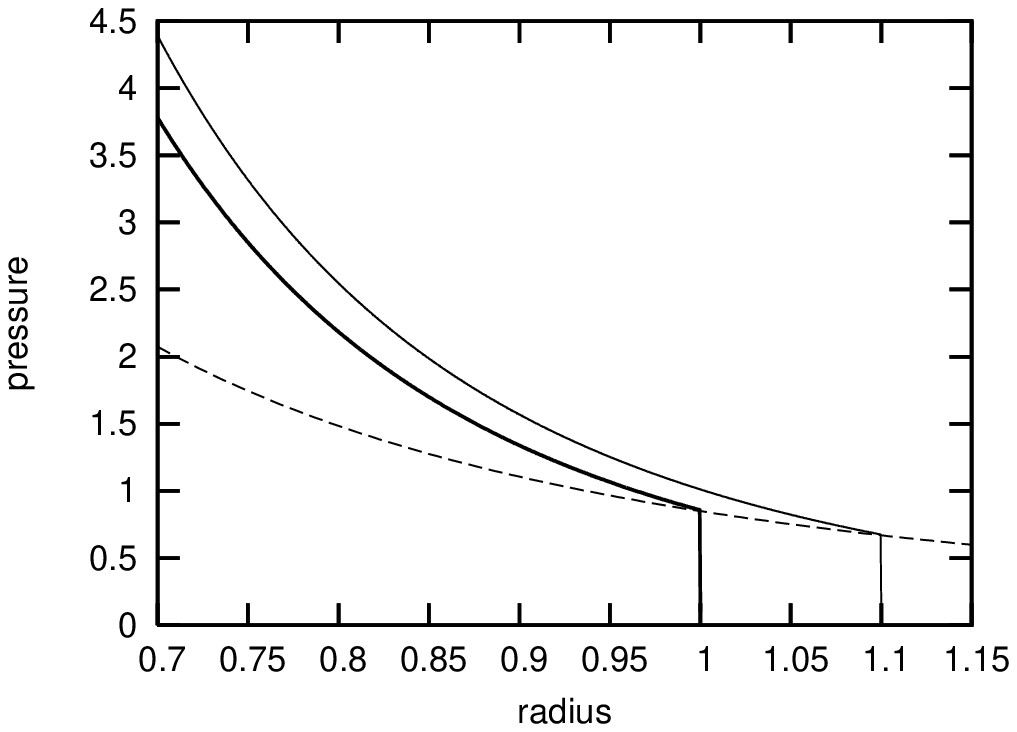}
\caption{The effect of changing the shock position on the post-shock pressure profile is illustrated 
here with two different equilibrium solutions.  The dashed line shows the immediate post-shock
pressure as a function of shock location.  Thus in each case the pressure begins on this line
at the position of the shock and varies approximately as $r^{-4}$ inward.  If the shock is displaced outward by a small
amount, from a radius of 1.0 to a radius of 1.1, the pressure of the inflowing gas 
behind the perturbed shock will start off at a lower value, but it will be higher at each radius below 1.0 than it is 
for the inflowing gas behind the original equilibrium shock. 
This positive feedback drives the growth of the acoustic waves, creating the 
SASI.}
\label{fig:presprofiles}
\end{center}
\end{figure}

The origin of the growth of this standing wave can be traced to the response of the post-shock
pressure to changes in the shock radius.  If the pressure in one hemisphere becomes slightly
higher than equilibrium, it will push the spherical accretion shock outwards.  Because the
preshock ram pressure drops with increasing radius (as $r^{-2.5}$), shown by the dashed line in 
Figure~\ref{fig:presprofiles}, the outward shock displacement leads to
a smaller pressure immediately behind the shock.  However, given that the postshock pressure 
increases with decreasing radius as $r^{-4}$, the postshock pressure for the perturbed shock
will be greater than the postshock pressure for the unperturbed shock at each radius below the
radius of  the original unperturbed shock. This is illustrated with the two cases shown in 
Figure~\ref{fig:presprofiles}. The postshock pressure immediately behind the shock for  the perturbed shock 
at radius 1.1 is lower than the postshock pressure immediately behind the unperturbed shock at 
radius 1.0 given the decrease in the preshock ram pressure given by the dashed line. However, 
at each radius below 1.0 the postshock pressure for the perturbed shock is higher than the postshock
pressure for the unperturbed shock.

There is an additional effect on the immediate post-shock
pressure due to the change in shock velocity.  As the shock is being pushed outward, the local
shock velocity is larger than for a stationary shock at that radius, leading to a slightly higher
post-shock pressure.  Note that the change in pressure due to shock velocity is out of phase
with respect to the change in pressure due to shock displacement, with the former peaking 
as the shock is moving outward, and the latter peaking at a phase $\pi/2$ later when the shock
has reached its maximum extent.  Nonetheless, both effects act to amplify the pressure variation
of the standing wave.  
For the observed frequencies of the $l=1$ mode of the SASI, the effect of changing shock velocity
is a few times smaller than the change in ram pressure due to shock displacement.

\begin{figure}[!hbtp]
\begin{center}
\includegraphics[width=6.5in]{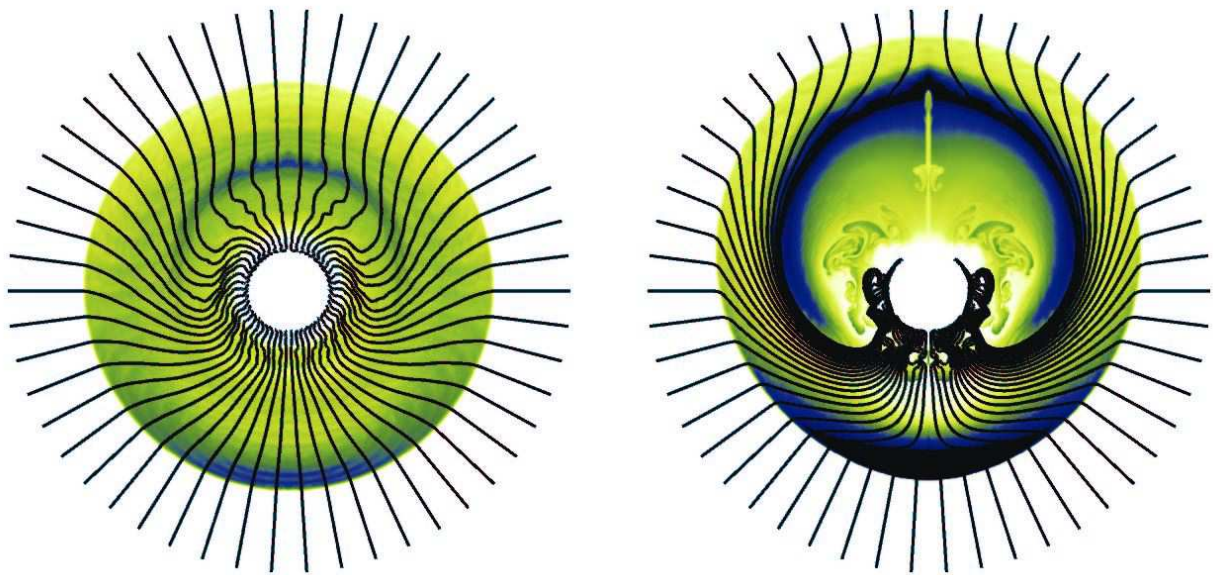}
\caption{The transition from the linear to the non-linear
regime is illustrated with these two images spanning one oscillation period.
The color shows positive (blue) and
negative (yellow/white) deviations to the equilibrium value of gas entropy, and the lines represent 
streamlines integrated along the instantaneous velocity field. The transition to the nonlinear regime
is largely characterized by the transition from radial to nonradial flow.  An
mpeg animation of this transition is available on line. 
}
\label{fig:transition}
\end{center}
\end{figure}

The linear phase of the SASI is characterized by a nearly spherical accretion shock and approximately
radial post-shock flow.  Once the amplitude of the standing pressure wave becomes large enough 
to significantly break the spherical symmetry of the accretion shock, the SASI enters the non-linear phase.
In this phase the radially infalling
gas above the shock strikes the shock surface at an oblique angle, generating strong, non-radial 
post-shock flow.  
This transition from the linear to non-linear phase is illustrated in Figure \ref{fig:transition}.  
The effect of a distorted shock on the post-shock flow is quite dramatic even for the relatively slight
changes in shock position shown in the second frame of Figure \ref{fig:transition}.  Although the accretion
shock is nearly spherical in this frame, it is significantly displaced upward.  As a result, the post-shock
flow is no longer radial, and in some regions is almost entirely tangential to the radial direction.  As a 
consequence of this non-radial flow, perturbations generated on one side of the shock can be advected
across the interior and over to the other side of the accretion cavity.  For example, the second frame
in Figure \ref{fig:transition} shows a shell of high-entropy gas (shown in blue) in the upper hemisphere
being advected around the central star and toward the lower hemisphere.

\section{Discussion and Conclusion}

The linear stability analysis presented here has illuminated the underlying
physical origin of the SASI instability first presented in
\citetalias{bmd03}. The SASI is not the result of a vortical--acoustic
feedback \citep{foglizzo02} seen in other contexts, as thought previously
(\citetalias{bmd03}). Rather, it is the result of a growing standing acoustic wave in the
spherical cavity bounded below by the surface of the compact object and
above by the accretion shock. This is our primary finding.

In addition, our techniques for exciting different modes in isolation, first
defined in \citetalias{bmd03} and refined here, and for defining and
extracting the growth rate of these modes in the linear regime, have
confirmed that the SASI is an $l=1$ instability, first proposed in
\citetalias{bmd03}.  This result provides a possible explanation for the large-scale
structure seen in recent two-dimensional supernova simulations performed on
numerical grids covering a full $\pi$ in angle \cite{jankaetal041}. Our results clearly
show that the $l=2$ mode is unstable, but we have not observed this mode becoming
nonlinear. Rather, the amplitude of the $l=1$ mode always overtakes that of the
$l=2$ mode during the linear regime.
In the two-dimensional case considered here, it is apparent that,
in the linear regime, power is transferred from the $l=2$ to the $l=1$ mode
with time.

Under near explosive conditions at the onset of a core collapse supernova (large 
neutrino heating region, thin neutrino cooling region), our results strongly suggest 
the SASI will develop. Moreover, the SASI has been confirmed in a two-dimensional
model by \cite{jankaetal042} that included neutrino transport and suppressed neutrino-driven 
convection in order to isolate, as we have done here, postshock flow induced by 
the SASI versus postshock flow induced by convection. And the development of 
an obvious $l=1$ mode in the explosion of an 11 M$_\odot$ progenitor in a simulation 
performed by the same group on a 180 degree angular grid, without any such suppression, 
is strong evidence for the SASI in a complete model that attains explosive conditions 
\cite{jankaetal041}. Moreover, given that this very same model did not explode when a 90 degree 
grid was used \cite{burasetal03}, one must consider that the SASI will play an important role in 
the explosion mechanism {\it per se}, as proposed in \cite{bmd03}, not just in defining gross characteristics of 
the explosion. 

Generally speaking, we have affirmed the discovery of the SASI in core
collapse supernovae \citetalias{bmd03} and supplanted our understanding of
its origin and development. Relatively small-amplitude perturbations,
whether they originate from inhomogeneities in the infalling gas, aspherical
pressure waves from the interior region, or perturbations in the postshock
velocity field, can excite perturbations in the standing accretion shock that
lead to vigorous turbulence and large-amplitude variations in the shape and
position of the shock front. In addition, we have confirmed the linear
stability analysis of \citetalias{hc92} for spherically symmetric modes,
providing a critical test of our numerical hydrodynamic algorithm in the
context of core-collapse supernovae.

\acknowledgments

This work is supported by a SciDAC grant from the U.S. DOE 
High Energy, Nuclear Physics, and Advanced Scientific
Computing Research Programs.
A.M. is supported at the Oak Ridge National Laboratory, managed by
UT-Battelle, LLC, for the U.S. Department of Energy under contract
DE-AC05-00OR22725.
We thank the Center for
Computational Sciences at ORNL for their generous support of computing 
resources.

\clearpage


\begin{thebibliography}{}

\bibitem[Blondin et al.(2003)]{bmd03}Blondin, J. M., Mezzacappa, A., \& DeMarino, C. 2003, ApJ, 584, 971
\bibitem[Buras et al.(2003)]{burasetal03}Buras, R., Rampp, M., Janka, H.-Th. \& Kifonids, K. 2003, PhRvL, 90, 1101
\bibitem[Burrows \& Goshy(1993)]{bg93}Burrows, A. \& Goshy, J. 1993, ApJ, 416, L75
\bibitem[Burrows et al.(1995)]{bhf95} Burrows, A., Hayes, J. \& Fryxell, B. A.1995, ApJ, 450, 830
\bibitem[Chevalier(1989)]{chev89}Chevalier, R. A. 1989, ApJ, 346, 847
\bibitem[Foglizzo(2002)]{foglizzo02}Foglizzo, T. 2002, A\&A, 392, 353
\bibitem[Fryer \& Warren(2004)]{fw04}Fryer, C. L. \& Warren, M. S. 2004, ApJ, 601, 391
\bibitem[Herant et al.(1992)]{hbc92}Herant, M., Benz, W. \& Colgate, S. A. 1992, ApJ, 395, 642
\bibitem[Herant et al.(1994)]{herantetal94}Herant, M., et al. 1994, ApJ, 435, 339
\bibitem[Houck \& Chevalier(1992)]{hc92}Houck, J. C. \& Chevalier, R. A. 1992, ApJ, 395, 592
\bibitem[Janka(2001)]{janka01}Janka, H.-T. 2001, A\&A, 368, 527
\bibitem[Janka et al. (2004) 1]{jankaetal041} Janka, H.-T., Buras, R., Kitaura Joyanes F.S., Marek, A. \& Rampp, M. 2004, astro-ph/0405289
\bibitem[Janka et al. (2004) 2]{jankaetal042} Janka, H.-T., Scheck, L., Kifonidis, K., M\"{u}ller, E. and Plewa, T. 2004, astro-ph/0408439
\bibitem[Janka \& M\"uller(1996)]{jm96} Janka, H.-Th. \& M\"uller, E. 1996, A\&A, 296, 167
\bibitem[Liebendoerfer et al.(2001)]{liebendoerferetal01}Liebendoerfer, M. et al. 2001, PhRvD, 63, 3004
\bibitem[Mezzacappa(2005)]{mezzacappa05}Mezzacappa, A. 2005, ARNPS, in press
\bibitem[Mezzacappa et al.(1998)]{mezzacappaetal98}Mezzacappa, A., et al. 1998, ApJ, 495, 911
\bibitem[Mezzacappa et al.(2001)]{mezzacappaetal01}Mezzacappa, A., et al. 2001, PhRvL, 86, 1935
\bibitem[Miller et al.(1993)]{mwm93} Miller, D. S., Wilson, J. R. \& Mayle, R. W. 1993, ApJ, 415, 278
\bibitem[Rampp \& Janka(2000)]{rj00}Rampp, M. \& Janka, H.-Th. 2000, ApJ, 539, L33


\end{thebibliography}
\end{document}